\documentclass[9pt,twoside]{rmaa-rho}
\RMxAAtemplatetype{\RMxAA} 

\usepackage[justification=justified,singlelinecheck=false]{caption}

\newcommand{\rxs}{1RXS~J174320.1-042953}

\vol{100}
\pages{1000-1025}
\thisyear{2026}
\doi{\href{https://doi.org/10.22201/ia.01851101p.20XX.XX.XX.XX}{https://doi.org/10.22201/ia.01851101p.20XX.XX.XX.XX}}

\title{1RXS J174320.1-042953: another polar with a red-shifted absorption component in  emission  line wings}

\author[1]{D. H. Gonz\'{a}lez-Buitrago \orcidlink{0000-0002-9280-1184}}
\author[1]{Ma. T. Garc\'ia-D\'iaz \orcidlink{0000-0002-9772-5555}}
\author[1]{S. Zharikov \orcidlink{0000-0003-2526-2683}}
\author[1]{G. Tovmassian \orcidlink{0000-0002-2953-7528}}
\author[1]{A. Castro \orcidlink{0000-0002-7832-5337}}
\author[2]{E.~J. Kotze}
\author[3]{Y. Krongold \orcidlink{0000-0001-6291-5239}}
\author[3]{C. A.Negrete \orcidlink{0000-0002-1656-827X}}

\affil[1]{Universidad Nacional Autónoma de México, Instituto de Astronomía, AP 106,  Ensenada 22800, BC, México}
\affil[2]{South African Astronomical Observatory, PO Box 9, Observatory 7935, Cape Town, South Africa}
\affil[3]{Universidad Nacional Autónoma de México, Instituto de Astronomía, AP 70-264, CDMX 04510, México}

\leadauthor{Gonzalez-Buitrago et al.}
\smalltitle{LaTex Macro RMxAA}

\corres{Diego Gonzalez-Buitrago}
\email{dgonzalez@astro.unam.mx}

\received{April 16th, 2024}
\accepted{\today}

\license{Texto de la licencia aquí}

\setbool{rho-abstract}{true} 
\setbool{rho-resumen}{true} 

\begin{abstract}
    We present the results of a spectroscopic and multi-band photometric study of the magnetic cataclysmic variable 1RXS J174320.1-042953. Multi-band photometry confirms a periodic modulation with an orbital period of $P_{\rm orb}=0.0864 \pm 0.0001$ days. Using the estimated secondary mass of $\sim0.15\,\mathrm{M_\odot}$ and, the X-ray-based white dwarf mass of $\sim0.75\,\mathrm{M_\odot}$, we obtain values of $q\sim0.2$ and a system inclination of $47^\circ\pm10^\circ$. The optical spectra of 1RXS J174320.1-042953 are dominated by strong, complex, asymmetric, and highly variable single-peaked emission lines of the Balmer series, \ion{He}{i}, and \ion{He}{ii}, which are characteristic of polar-type magnetic cataclysmic variables. Doppler tomography reveals no evidence of a Keplerian accretion disc, supporting the classification of the system as a polar. The H$\beta$ and \ion{He}{ii} emission lines can be decomposed into at least two distinct components. The low-velocity components have semi-amplitudes of approximately 170 km s$^{-1}$ (H$\beta$) and 147 km s$^{-1}$ (\ion{He}{ii}), while the high-velocity components reach about 317 km s$^{-1}$ and 460 km s$^{-1}$, respectively. The low-velocity component is likely associated with the irradiated side of the secondary star facing the white dwarf and/or the vicinity of the L$_1$ point, whereas the high-velocity component is related to the accretion-stream structure. A redshifted absorption component in the emission-line wings is detected at orbital phases near $\phi \approx 0.0$, reaching velocities up to $\sim1400$ km s$^{-1}$ and likely produced by accretion-stream material crossing the line of sight.
\end{abstract}

\keywords{binaries, Cataclysmic Variables, 1RXS~J174320.1-04295}

\begin{resumen}
    Presentamos los resultados de un estudio espectroscópico y fotométrico multibanda de la variable cataclísmica magnética 1RXS J174320.1-042953. La fotometría multibanda confirma una modulación periódica con un periodo orbital de $P_{\rm orb}=0.0864 \pm 0.0001$ días. Utilizando la masa estimada de la estrella secundaria de $\sim0.15\,\mathrm{M_\odot}$ y, la masa de la enana blanca de $\sim0.75\,\mathrm{M_\odot}$ obtenida a partir de rayos-X, obtenemos valores de $q\sim0.2$ y una inclinación del sistema de $47^\circ \pm 10^\circ$. Los espectros ópticos de 1RXS J174320.1-042953 están dominados por líneas de emisión intensas, complejas, asimétricas y altamente variables de la serie de Balmer, \ion{He}{i} y \ion{He}{ii}, características de variables cataclísmicas magnéticas de tipo polar. La tomografía Doppler no revela evidencia de un disco de acreción kepleriano, lo que respalda la clasificación del sistema como un polar. Las líneas de emisión H$\beta$ y \ion{He}{ii} pueden descomponerse en al menos dos componentes distintas. Las componentes de baja velocidad presentan semiamplitudes de aproximadamente 170 km s$^{-1}$ (H$\beta$) y 147 km s$^{-1}$ (\ion{He}{ii}), mientras que las componentes de alta velocidad alcanzan valores cercanos a 317 km s$^{-1}$ y 460 km s$^{-1}$, respectivamente. La componente de baja velocidad está probablemente asociada con el lado irradiado de la estrella secundaria que enfrenta a la enana blanca y/o con la vecindad del punto L$_1$, mientras que la componente de alta velocidad está relacionada con la estructura del flujo de acreción. Se detecta una componente de absorción desplazada al rojo en las alas de las líneas de emisión en fases orbitales cercanas a $\phi \approx 0.0$, alcanzando velocidades de hasta $\sim1400$ km s$^{-1}$ y probablemente producida por material del flujo de acreción que cruza la línea de visión.
\end{resumen}

\begin{document}

\nolinenumbers

\maketitle
\pagestyle{fancy}\thispagestyle{firststyle}

\section{Introduction}
\label{sec:intro}

Cataclysmic variables (CVs) are semi-detached close binary systems composed of a white dwarf (WD) accreting material from a late main-sequence star or a brown dwarf. The companion star fills its Roche lobes and transfers material to the WD through the  Lagrange point $L_{1}$ \citep[for a review, see][]{Warner:1995}. The strength of the magnetic field of the WD plays a pivotal role in the accretion  process in CVs. If the magnetic field is weak \citep[$B < 10^7~ {\rm G}$;][]{Martino:2004}, transferred material spirals toward the WD, generating an accretion disk around it.   For systems with moderate magnetic fields ($B \lesssim 10^7~ {\rm G}$),known as intermediate polars transferred material can form a partial disk, with its inner part disrupted into accretion curtains channeling material to the white dwarf's magnetic poles. In such systems, the spin ($P_{\rm spin}$)  of the WD is asynchronous with the orbital period ($P_{\rm orb}$), usually yielding $P_{\rm spin}/P_{\rm orb}\sim$0.01--0.6 \citep{Norton:2008}. 
If the magnetic field is sufficiently strong ($B > 10^7,G$), accretion disk 
formation is impeded. Instead, the magnetic field governs the flow of transferred 
material, directing it along magnetic lines to the white dwarf's poles.  
This process leads to the material free-falling onto the white dwarf's surface, 
resulting in a strong shock. This shock generates X-ray, UV, and optical emissions, predominantly governed by cooling processes such as Bremsstrahlung and cyclotron emission \citep{Cropper:1990, Wickramasinghe:2000}. The broad cyclotron emission  lines can be observed as a distinctive hump in the optical and infrared spectra. 

Additionally, strong magnetic-field systems, known as AM Her-type or polars or magnetic CVs (mCVs), are characterized by prominent spectral features, including strong Balmer emission lines, \ion{He}{I}, \ion{He}{II}, and \ion{C}{III}/\ion{N}{III} emission blend at $\sim$4650\AA. These features, primarily originating in the accretion column, exhibit single peaked and asymmetric profiles \citep{Warner:1995}. In most cases the rotational periods of the WDs are synchronized with the system's orbital periods in polar systems \citep{Cropper:1990}.
More recent studies estimate that about 35\% of CVs are magnetic systems, with polars representing approximately 29--30\% of the CV population \citep{Pala2020,Rodriguez2025}.

\begin{figure*}
\centering
\begin{tikzpicture}
  \node[anchor=north west,inner sep=0] (image1) at (.3,0)
  {\includegraphics[width=8.6in, bb=10 10 1200 330, clip]{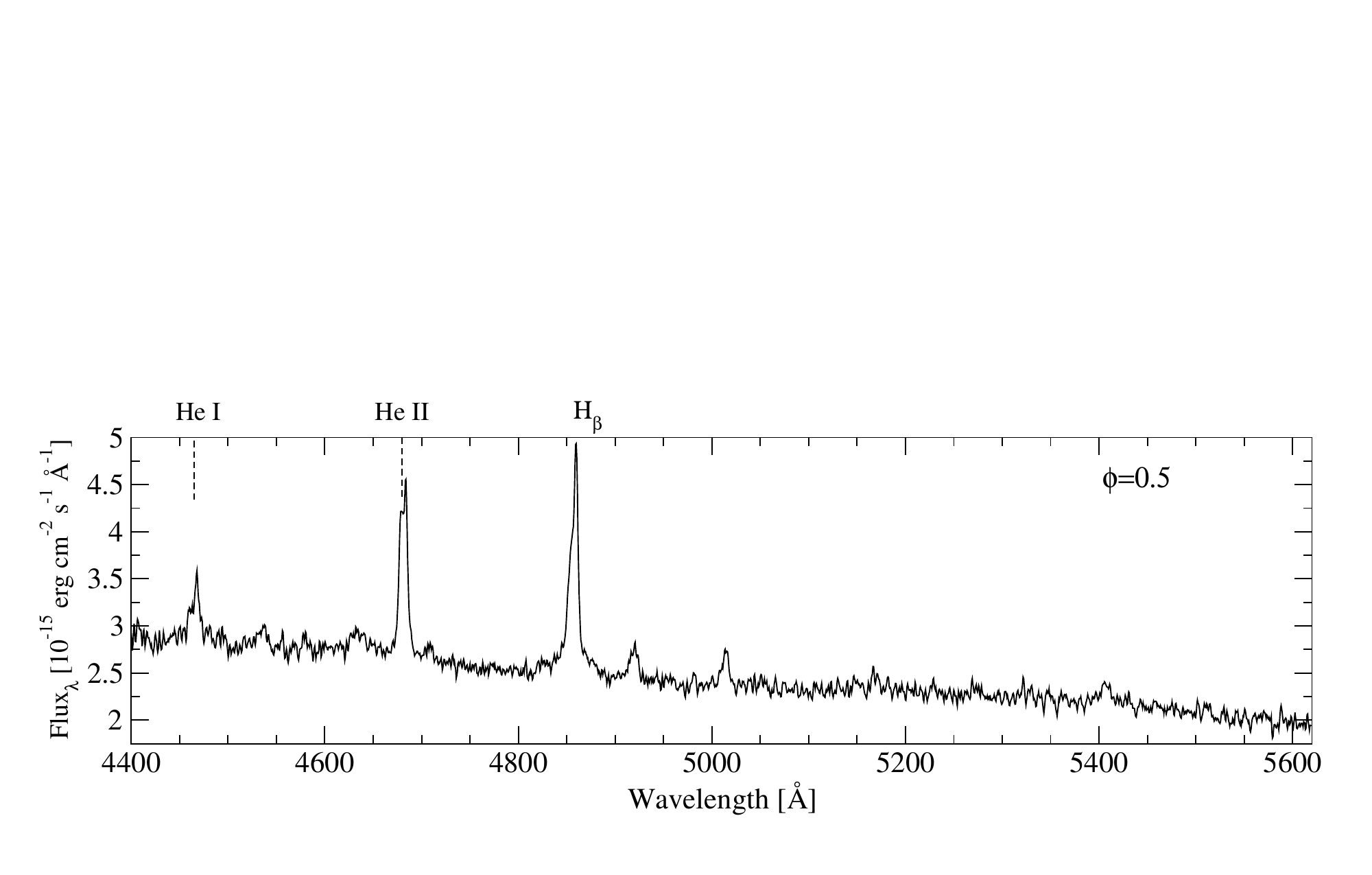}};
  \pgfmathsetmacro{\yShift}{-5.5}
  \node[anchor=north west,inner sep=0] (image2) at (0.0,\yShift)
  {\includegraphics[width=7.4in, bb=10 100 1200 380, clip]{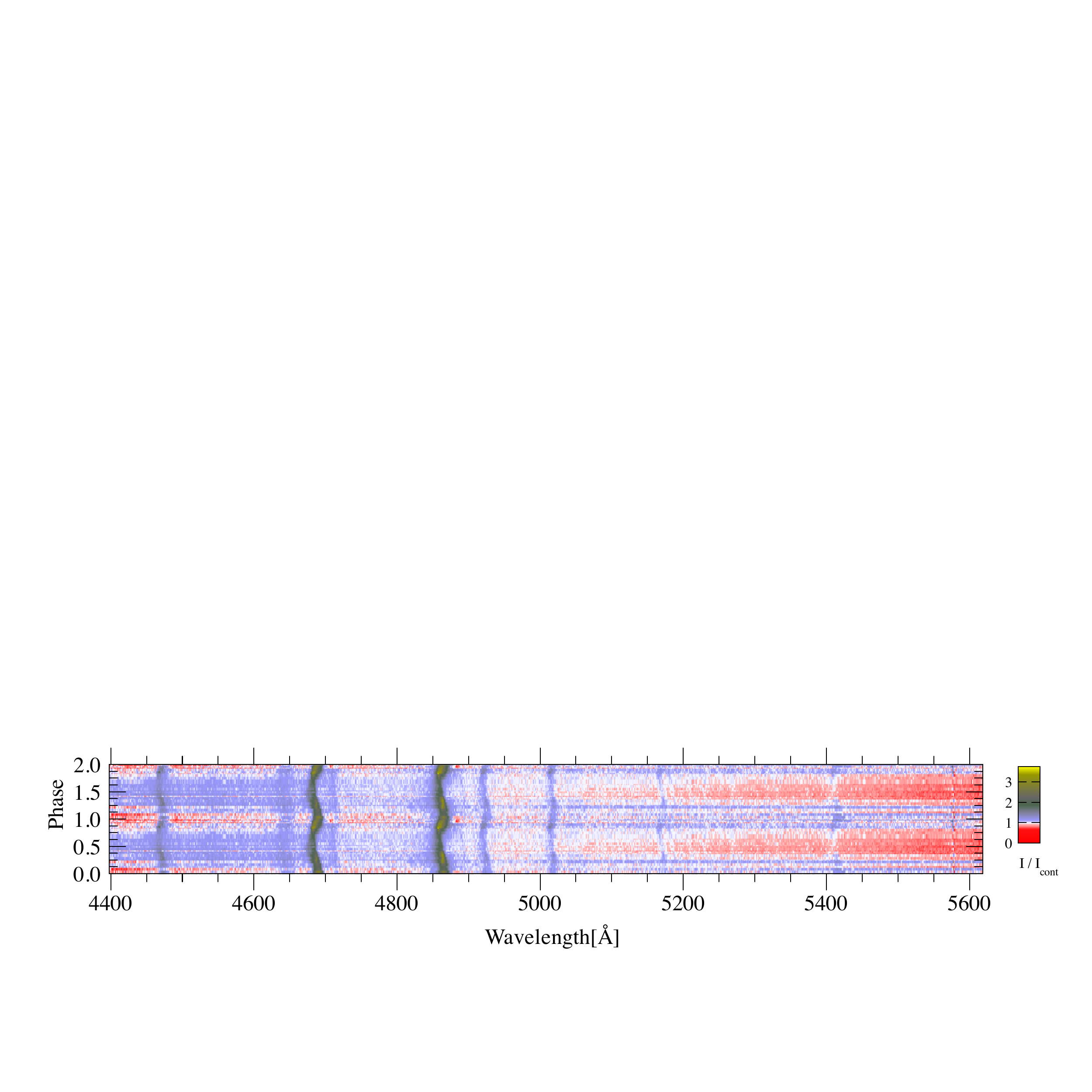}};
  \pgfmathsetmacro{\yShift}{-8.5}
  \node[anchor=north west,inner sep=0] (image3) at (.3,\yShift)
  {\includegraphics[width=8.6in, bb=10 10 1200 330, clip]{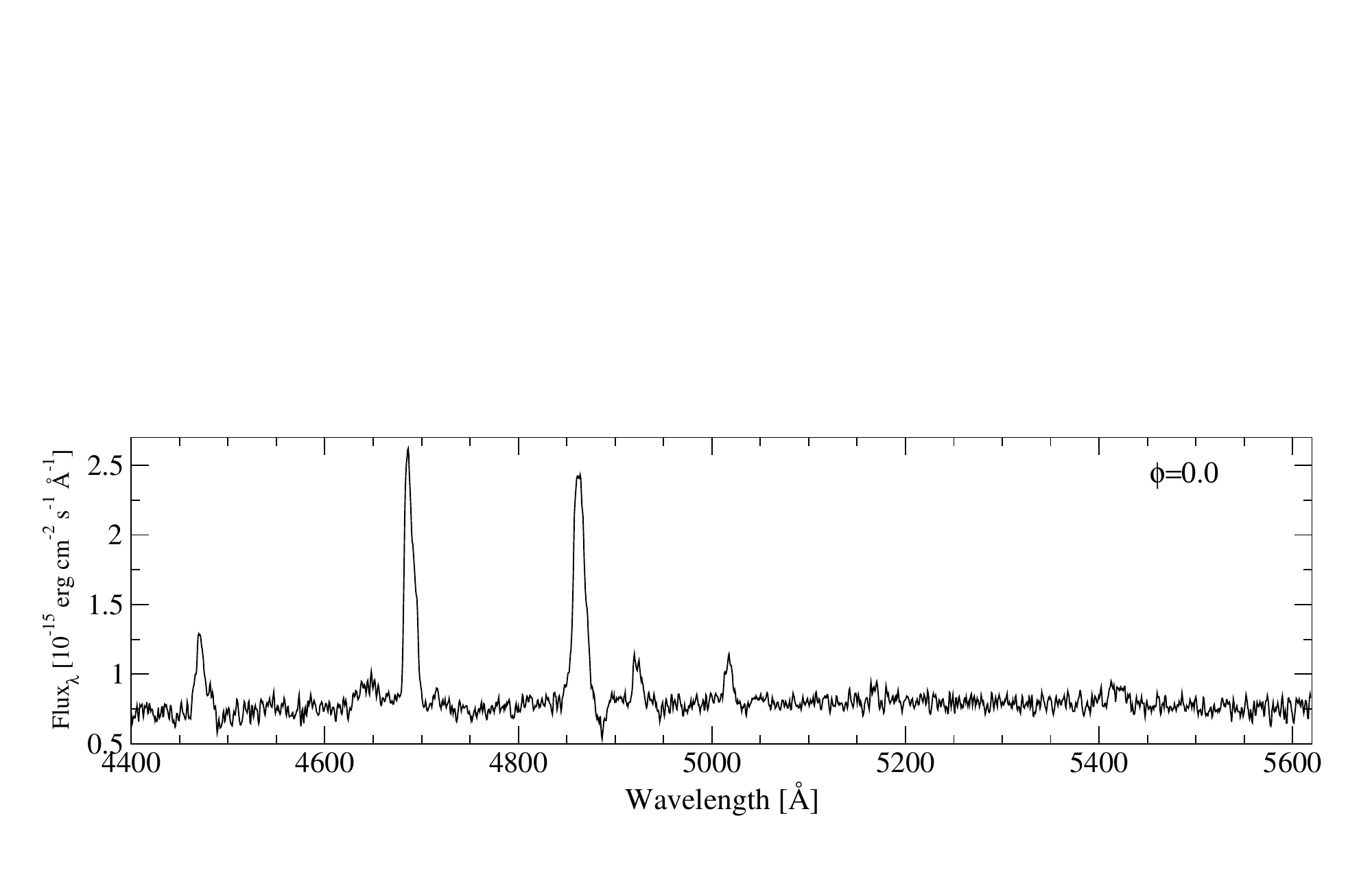}};
  \begin{scope}[
      shift={(image3.south west)},
      x={(image3.south east)},
      y={(image3.north west)}
  ]
      \def\lambdaTarget{4820}
      \def\lambdaMin{4400}
      \def\lambdaMax{5600}
      \pgfmathsetmacro{\xpos}{(\lambdaTarget-\lambdaMin)/(\lambdaMax-\lambdaMin)}
      \draw[->, red, line width=1pt]
          (0.52,0.80) -- (\xpos,0.35);
      \node[red, font=\small, anchor=south]
          at (0.52,0.82) {abs.\ line};
  \end{scope}
\end{tikzpicture}
\caption{\textbf{Upper panel:} Average spectrum of \rxs{} at orbital phase $\phi=0.5$.  \textbf{Middle panel:} Trailed spectra covering the wavelength range 4400--5600\,\AA.  \textbf{Lower panel:} Average spectrum of \rxs{} at $\phi=0.0$, where an absorption feature is visible near the H$\beta$ emission line.}
\label{fig:spec_phase}
\end{figure*}

\rxs{} was identified as a cataclysmic variable by \citet{Denisenko:2011}, who suggested its classification as an intermediate polar due to the presence of small fluctuations 
in the photometric data. 
\citet{2012PZ.....32....3D} found that the orbital 
period of the system is 124.7 min~/~0.0866(3) days, 
placing the system at the lower edge of the CV period 
gap \citep[and references therein]{Knigge:2011}.
Additionally, authors indicated the presence of shorter period variations superimposed on the main orbital curve, which they speculated could signify characteristics of an intermediate polar with a rapidly rotating WD. Subsequently, \citet{Oliveira:2017A} performed optical spectroscopy of  \rxs{} and found that its spectra demonstrate characteristics inherent to a polar showing intense single peaked emission Balmer, \ion{He}{I}, and \ion{He}{II} ($\lambda 4686$\AA) lines. In recent XMM-Newton X-ray and optical study of  \rxs{}, conducted by \citet{Rawat:2023}, the object's high and low luminosity states were observed. Authors found that the mass-transfer rate in the system is $\sim 5\times 10^{-12}\,\,{\rm M}_{\odot}\,{\rm yr}^{-1}$,  the bolometric luminosity of  $\sim4 \times 10^{31}~{\rm erg\,s^{-1}}$, and they estimated the white dwarf mass of 0.75$^{+0.16}_{-0.11}$ ${\rm M}_\odot$ from modeling the X-ray emission of the post-shock region.
 The Gaia distance to the object is 214$\pm$2 pc 
 \citep{2021AJ....161..147B} and corresponding interstellar 
 absorption 
 $E(g-r)$$=0.35_{-0.09}^{+0.03}$ \citep{2019ApJ...887...93G}.

In this paper we report new spectroscopic and  multi-band photometric observations of \rxs{}   to determine system parameters and its accretion stream structure. In  Section \ref{sec:observation}, we describe observations and data reduction. Section \ref{sec:analisis} contains our analyses of the light curves, determination of the structure of emission lines and their radial velocities (RVs), estimation of binary system parameters, and result of Doppler mapping.
The discussion and conclusions are presented in Sections~\ref{sec:discusion} and~\ref{sec:conclusion}.


\section{Observations and data reduction}
\label{sec:observation}

\subsection{Spectroscopy}
\label{sec:spec}

We conducted time-resolved spectroscopy of \rxs{} using Boller \& Chivens spectrograph attached 2.1m telescope of Observatorio Astron\'omico Nacional de  San Pedro M\'artir, Mexico (OAN-SPM\footnote{https://www.astrossp.unam.mx}). The object was observed during four nights of 2015 June. Most of the data was obtained using  1200 \rm{l/mm} grating that provides 2.1\AA\ spectral resolution for the range of  4400 to 5600\AA\ (Fig.~\ref{fig:spec_phase}). In addition, we got three spectra with  6.4\AA\  spectral resolution covering from 4000 to 7000\AA\ range, obtained with the same Boller \& Chivens spectrograph at the 2.1-m telescope at OAN-SPM. Alongside the \rxs{} observations, spectrophotometric standards from \citet{Oke:1990}  were observed, and the Cu-Ne-Arg lamp was used to calibrate the wavelength.
The standard long-slit spectroscopic data reduction was performed using corresponding {\sc iraf}
tasks. The log of the observations is given in Table~\ref{tab:log:spec}.

\begin{table}
\caption{Log of spectroscopic observations .}
\begin{tabular}{l|ccc} 
\hline \\[0.1pt]
Date  & Exp.Time &    Range &   N$^1$/hours$^2$     \\
(dd/mm/yr)       &   (s)   & (\AA)&           \\   \hline \\
11/06/15   &  900   & 4400--5620 &  12/3.14     \\
12/06/15   &  900    & 4400--5620 &  12/3.14      \\
13/06/15   &  900    & 4400--5620 &  12/3.00     \\
13/06/15   &  1200   & 4000--7000 &  3/1.00     \\
15/06/15   &  900    & 4400--5620 &  10/3.00        \\
\hline
\end{tabular}

$^1$-Number of spectra; $^2$-duration
\label{tab:log:spec}
\end{table}

\begin{table}
\caption{Log of photometric observations.}
\begin{tabular}{l|ccl} 
\hline  \\
Date        & Exposure        &   Filter  &  hours \\
(dd/mm/yr)  &     (s)         &            &  \\ [1pt]   \hline \\[0.1pt]
12/08/17$^*$     & 120/60/50 & B/V/R &  4.2 \\
13/08/17$^*$     & 120/60/50 & B/V/R &  4.3 \\
14/08/17$^*$     & 120/60/50 & B/V/R &  4.0 \\
18/04/24$^\star$ & 7  & g$^\prime$/r$^\prime$ /z$^\prime$   &   2.0 \\
[1pt]
\hline
\end{tabular}

$^*$-0.84m/MEXMAN; $^\star$-2.1m/OPTICAM
\label{tab:log}
\end{table}

\subsection{Photometry}
Multi-bands Johnson-Cousins BVR$_\mathrm{c}$I$_\mathrm{c}$  and the g$^\prime$r$^\prime$z$^\prime$ SDSS photometric observations of \rxs{} were obtained using the 0.84-m and 2.1-m telescopes of  OAN-SPM in the direct CCD imaging mode ({\sc mexman}) and the ultra-fast {\sc opticam}  camera~\citep{Castro2019, Castro2024}, respectively. 
Standard data reduction was performed using {\sc iraf}   tasks and  {\sc python} pipeline. 
We estimate our magnitude errors in the range of 0.01 to 0.05 magnitudes. 
The log of photometric observations is given in Table \ref{tab:log}. 

\section{Analysis}
\label{sec:analisis}
\subsection{Emission lines}
\label{sec:RadialVel}

\begin{figure*}
    \includegraphics[width=0.95\textwidth]{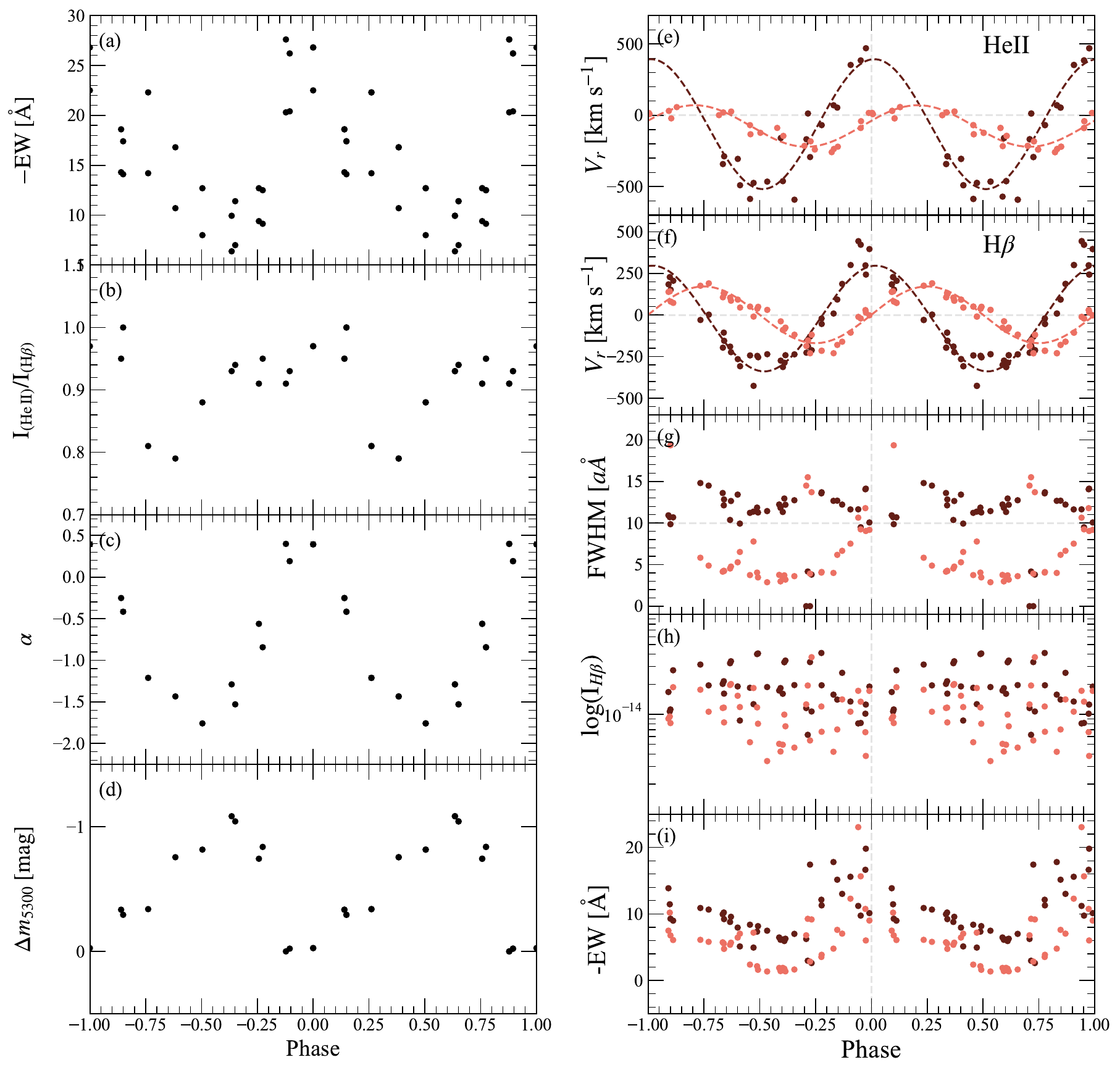}
  \caption{Orbital-phase dependence of the spectroscopic properties of H$\beta$ and He\,\textsc{ii}. Panels (a)-(d) show the quantities derived from the integrated spectral properties: (a) equivalent width of the H$\beta$ emission line, (b) peak intensity ratio $\mathrm{I}_{\mathrm{He\,II}}/\mathrm{I}_{\mathrm{H}\beta}$, (c) variation of the continuum slope $\alpha$, and (d) continuum flux expressed in magnitudes at $\lambda_c = 5300$\,\AA\ ($\Delta\lambda = 100$\,\AA). Panels (e)-(i) present the parameters obtained from the two-Gaussian decomposition of the emission lines: (e) radial velocity curve of the Gaussian components fitted to He\,\textsc{ii} $\lambda4686$, (f) radial velocity curve of the Gaussian components fitted to H$\beta$, (g) full width at half-maximum (FWHM) of H$\beta$ for both components, (h) logarithmic H$\beta$ line flux computed as $\mathrm{I}_{H\beta} = |{\rm EW}|\times F_{\rm cont}$, and (i) equivalent width of H$\beta$ for the two Gaussian components. For clarity, the orbital phase is repeated over the interval $[-1,1]$, and dashed curves indicate sinusoidal fits to the radial velocity measurements.}
  \label{fig:vr2}
\end{figure*}

\begin{figure}
\includegraphics[height=2.3in,bb = 250 100 850 1000, clip=]{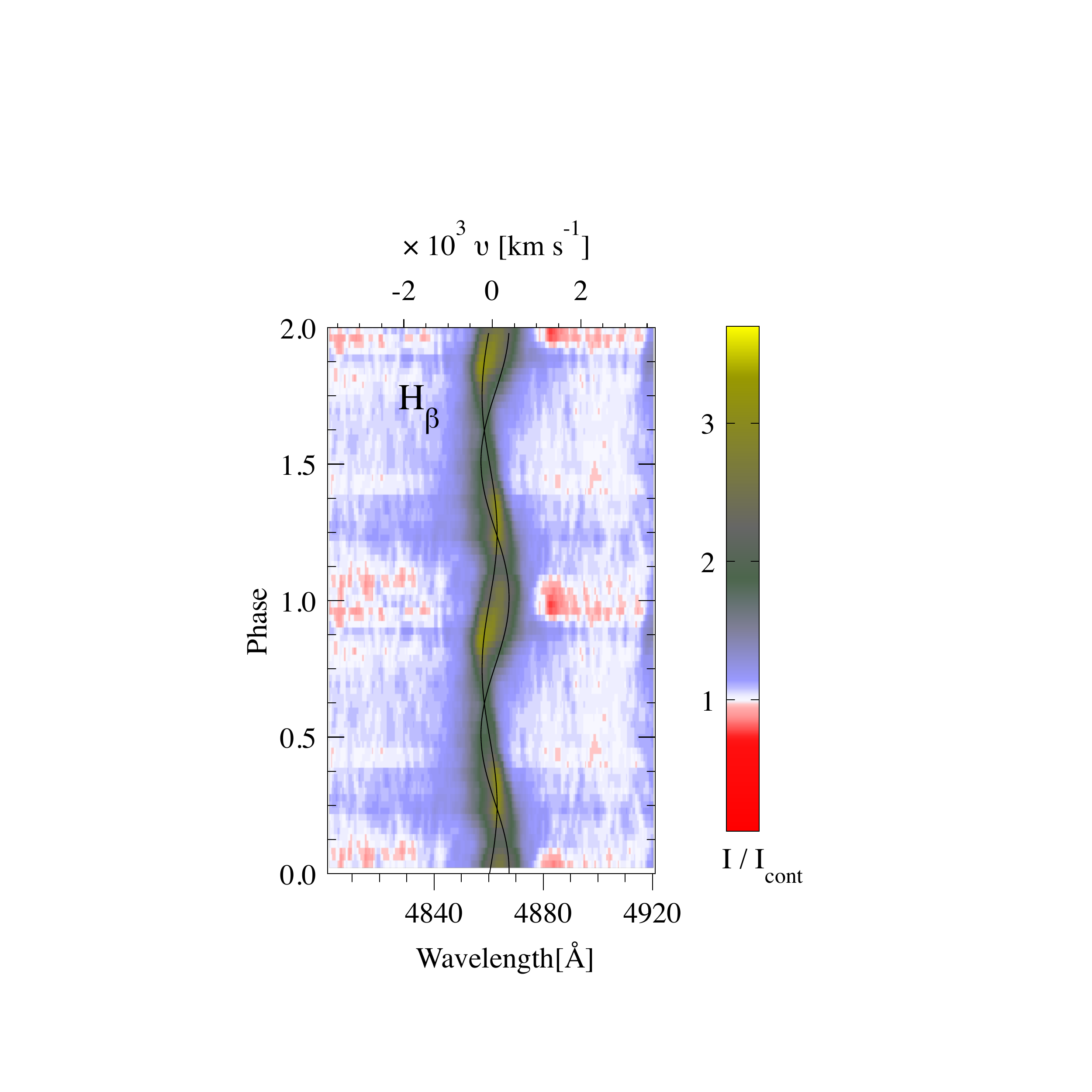}
\includegraphics[height=2.3in ,bb = 250 100 850 1000, clip=]{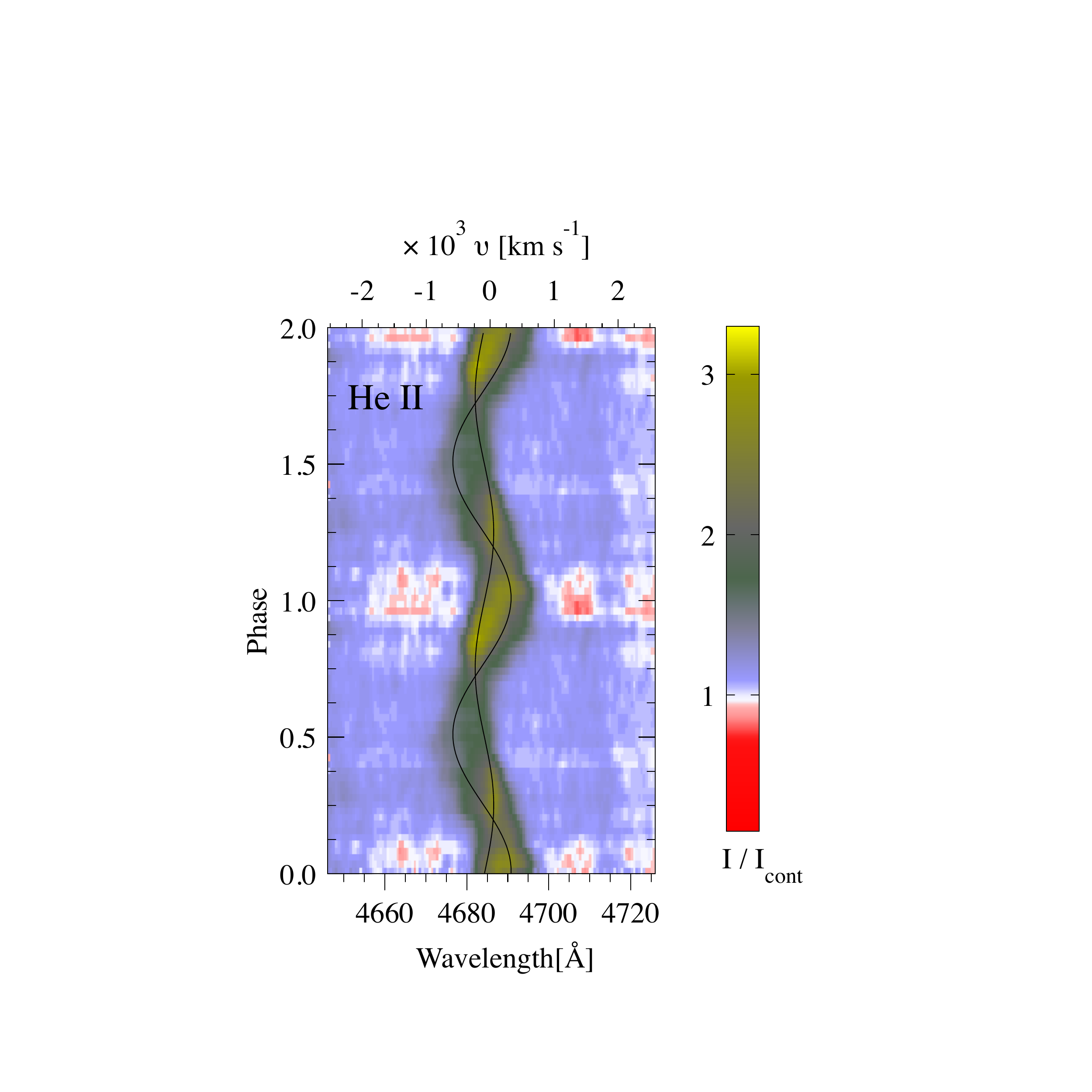}
\caption{Trailed spectra of the emission lines  H$\beta$ (left panel) and \ion{He}{ii} (right panel). The continuous black line in each trailed spectrum corresponds to the orbital modulation of the radial velocity for each line, achieved through the fitting of two Gaussians, as described in Section~\ref{sec:RadialVel}.}
\label{fig:trailed}
\end{figure}

\begin{figure}
\centering
  \includegraphics[width=0.5\textwidth,bb = 0 0 1000 340, clip=]{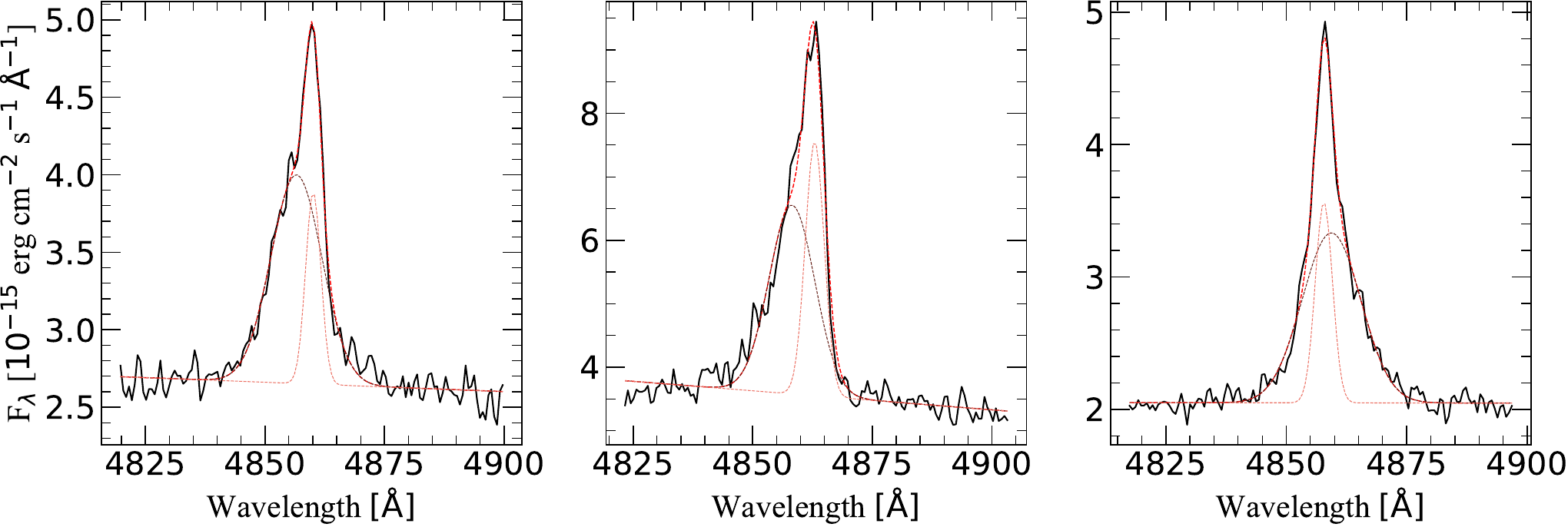}
  \caption{The figure shows the H$\beta$ emission line from three spectra taken at different phases. The black dotted lines correspond to the two Gaussians fitted to the emission line, and the continuous red line represents the sum of the two Gaussians.}
  \label{fig:specfit}
\end{figure}

The general characteristics of \rxs{} spectrum were described by \citet{Rawat:2023}.
In Fig.~\ref{fig:spec_phase}, top and bottom panels,  we present the spectra  of \rxs{} obtained by us in 2015 June, 11 in the maximum and minimum its optical brightness, respectively. The spectrum reveals intense emission lines from the Balmer and \ion{He}{II} ($\lambda$4686 \AA), as well as \ion{He}{I} at 4471, 4921, 5015, and 5876 \AA. The middle panel of Fig.~\ref{fig:spec_phase} presents phase-resolved (trailed) spectra of \rxs{} normalized to the flux in the V- band.
There is strong variation of the continuum flux, the  spectrum slope $\alpha$ ($\propto\lambda^\alpha$), and equivalent width (EW) of emission lines along the orbital phases (Fig~\ref{fig:vr2}-left panels). The emission lines (Figs.~\ref{fig:spec_phase},\ref{fig:trailed}, and \ref{fig:specfit}) have complicated, highly variable single-peaked  asymmetric profiles, characteristic of most mCVs.  
The line's patterns are identical to that observed in  many other systems \citep{Ok2024}. Following the custom for other AM Her systems, the H$_\beta$ and \ion{He}{II} $\lambda$4686\AA\ lines were deconvolved with two Gaussians. The examples of such fits of H$\beta$ profiles from some spectra are illustrated in Fig.~\ref{fig:specfit}.  The results of de convolution are presented 
in Table~\ref{tab:vel}, where the low-velocity component is marked by the letter ``L" and the high one by ``H", and at Fig.~\ref{fig:vr2} (right panel) and Fig.~\ref{fig:trailed}   for H$\beta$ and \ion{He}{II} lines. The low-velocity component has  amplitudes of  171$\pm7$~km~$s^{-1}$  and 147$\pm10$~km~$s^{-1}$ for for H$\beta$ and \ion{He}{II}, respectively. The systemic velocity ($\gamma = 1 \pm 4$~km~$s^{-1}$) is consistent  with the rest frame within uncertainties. The phase zero was selected that it corresponds  the switch of RV  of the low velocity component from negative to positive values: 
\begin{equation}
HJD = 2457187.7231 + 0^d.0864 \times E,
\label{eq:ephemerides}
\end{equation}

The high-velocity component is shifted  $\approx$0.25 orbital phase  relative to the low-velocity component. Its  amplitude reach 317$\pm11$~km~$s^{-1}$ for $H\beta$ and $\approx$460~km~$s^{-1}$ for \ion{He}{II}, respectively.  The low-velocity component is more intense but, generally, narrower its FWHT about two times less than the  high-velocity one.

\begin{table}
\caption{Spectroscopy parameters.}
\label{tab:vel}
\begin{tabular}{@{}ccccc}
\hline\hline
Line & component & $\gamma$ & K & $\phi$ \\
     &           & ($\mathrm{km~s^{-1}}$) & ($\mathrm{km~s^{-1}}$) &  \\
\hline
H$\beta$ & L &  1$\pm$4 & 171$\pm7$ & $0.00\pm$0.03 \\
         & H & -21$\pm$8 & -317$\pm11$ & $0.42\pm$0.42 \\
\ion{He}{ii} & L & -75$\pm$7 & 147$\pm10$ & $0.11\pm$0.05 \\
         & H & -62$\pm$25 & -460$\pm33$ & $0.58\pm$0.07 \\
\hline
\end{tabular}

\vspace{2mm}
\noindent {\it Note.} $RV(t) = \gamma + K \sin \left( \frac{2\pi t}{P} + \phi \right)$,
where $\gamma$ is the systemic velocity, $K$ is the semi-amplitude, $P$ is the orbital period, and $t$ is time.
\end{table}

\subsection{Absorptions} 

There is a remarkable feature in the spectra of \rxs{}, consisting of a redshifted absorption component detected in the wings of the emission lines at specific orbital phases. 
An example of this feature is shown in Fig.~\ref{fig:spec_phase} (lower panel), and it is also clearly visible in the trailed spectra (Fig.~\ref{fig:trailed}), where it appears as a transient structure confined to a limited range of phases. 
The best measurement of the absorption feature in the H$\beta$ line yields $\lambda = 4884.2 \pm 1.5$~\AA. Assuming that this feature originates from H$\beta$, we derive a redshifted velocity of $\sim 1410 \pm 94$ km s$^{-1}$. Such high velocities are consistent with material moving within the accretion flow in magnetic cataclysmic variables, and suggest that the absorbing region is located relatively close to the white dwarf.
The phase-dependent appearance of this feature indicates that the absorbing material intersects the line of sight only during a restricted interval of the orbital cycle, pointing to a geometric origin. In particular, this behavior is consistent with material in the accretion stream temporarily obscuring the emission region as it is guided along magnetic field lines toward the white dwarf.

The origin of this absorption component and its connection to the accretion flow are discussed in Section~\ref{sec:discusion}.


\subsection{Light curve and orbital  modulation}
\label{sec:lc}

\begin{figure*}
\centering
\includegraphics[width=1.0\textwidth]{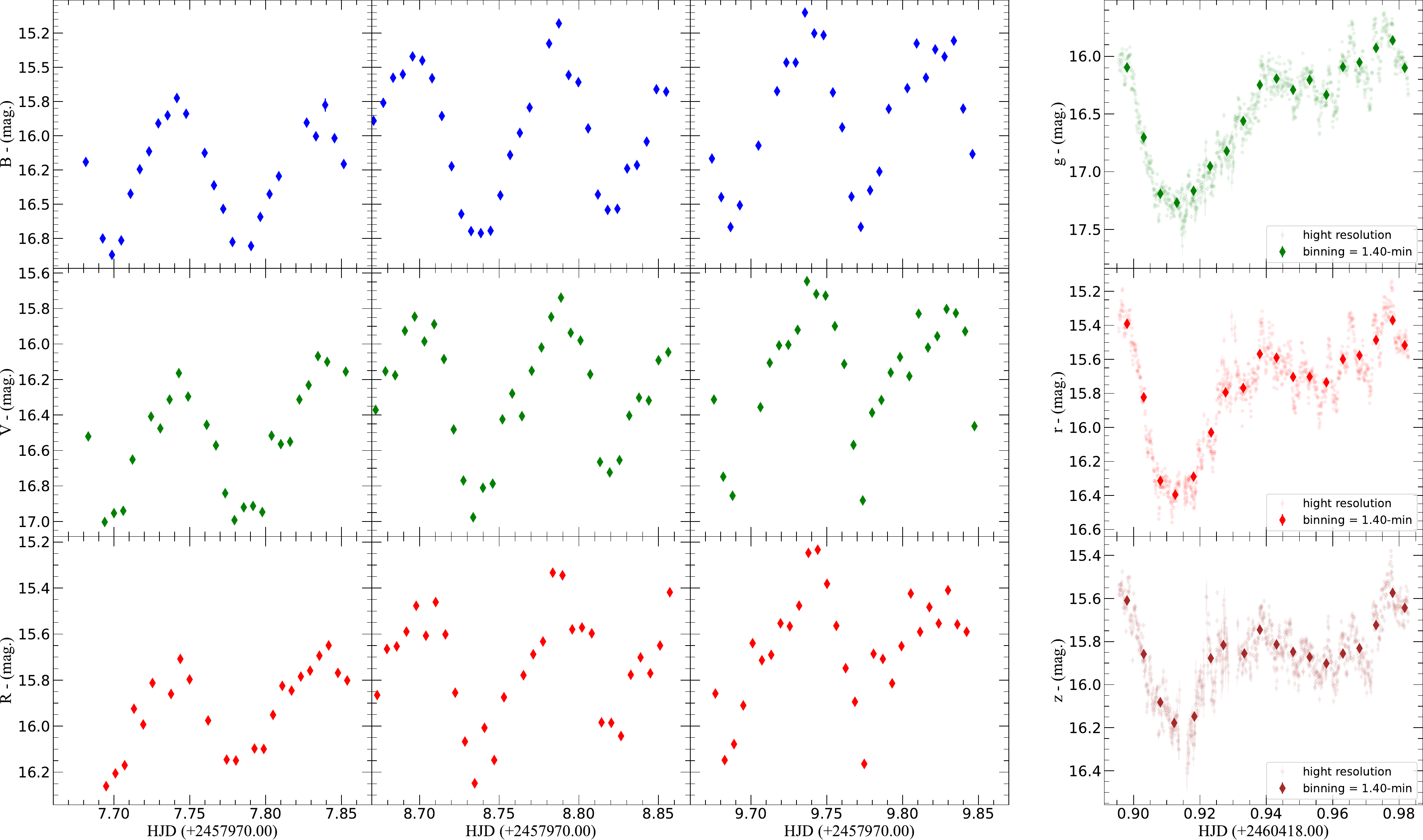}
\caption{left) The figure shows the light curve of \rxs\ in the BVR$_c$ bands bands. Each frame corresponds to one night of observation. 
right) The high-time resolution OPTICAM $g'r'z'$-bands of \rxs\ obtained in 2024 April, 18.}
\label{fig:UBVRI}
\end{figure*}

\begin{figure*}
\centering
\setlength{\unitlength}{1mm}
\resizebox{10cm}{!}{
\begin{picture}(100,145)(0,0)
\put (-15,97)  {\includegraphics[width=5.1cm, bb = 1000 500 3400 2200, clip=]{Fig6a.pdf}}
 \put (39,97)  {\includegraphics[width=5.1cm, bb = 2000 500 4000 2500, clip=]{Fig6b.pdf}}
 \put (72,107)  {\includegraphics[width=5.1cm, bb = 2000 1000 4000 2500, clip=]{Fig6c.pdf}}
 \put (-40,0)  {\includegraphics[trim=0.0cm 0.0cm 0.0cm 0.0cm, clip, width=\textwidth]{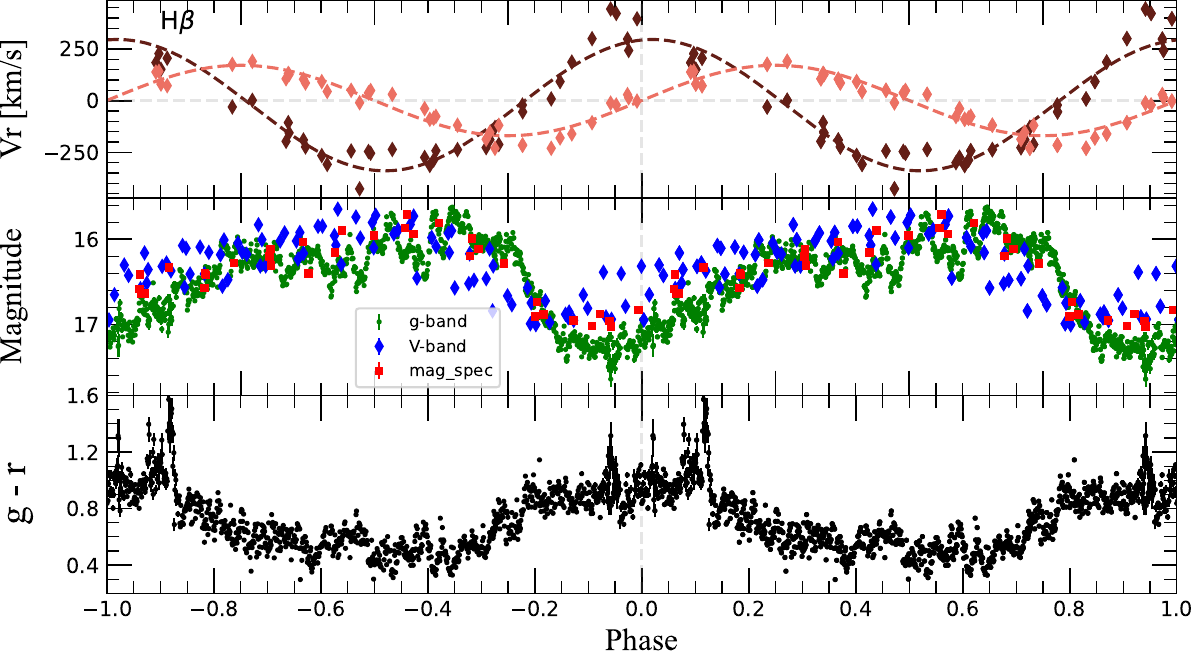}}
 \put (55.25,110) {\vector(0,-1){10}}
 \put (21,115) {\vector(3,-4){10}}
 \put (89,115) {\vector(-3,-4){10}}
 \put (55.25,100) {\line(0,-1){3}}
 \put (31.45,100) {\line(0,-1){3}}
 \put (79.1,100) {\line(0,-1){3}}
 \end{picture}
 }
    \caption{The top panel shows the radial velocity curves for the H$\beta$ line, fitted with two Gaussian components. Each Gaussian is plotted using its corresponding color.
    The middle panel presents the light curves for the $g$ band (green), $V$ band (blue), and the spectro-photometry light curve (red), which was obtained by integrating the flux around a central wavelength of 5477\AA\, with a bandwidth of 200\AA.
    The bottom panel displays the color index curve for the $g - r$ bands.}
    \label{fig:color}
\end{figure*}

Figure~\ref{fig:UBVRI} shows the light curves of \rxs{} in BVR$_c$ (left panel) and $g'r'z'$ bands (right panel) for each night of observations. In the BVR$_c$ light curves, a sine-like modulation is observed. The magnitude amplitude varies from night to night, similar to what was reported by \citet{Rawat:2023}. The FFT period search\footnote{PERIOD04 code \citep{2005CoAst.146...53L}} shows a prominent peak in the power spectrum at $f=11.5660\,\mathrm{d^{-1}}$, corresponding to $P_{\rm orb}=0.0864\pm0.0001$ days, coinciding with the period reported by \citet{Denisenko:2011} and \citet{Rawat:2023} within its uncertainties. The periodogram also shows the expected daily alias structure at $\pm1\,\mathrm{d^{-1}}$, although these peaks are significantly weaker than the main frequency.
The high-time-resolution OPTICAM observations are consistent with the lower-resolution ones and reveal a more complex structure in the object's light curve. There is a sharp drop in the object brightness of up to 1.8 magnitudes, with a duration of approximately $\sim45$ minutes. After that, a slower rise in brightness leads to a plateau close to the initial brightness level. The light curves also show substantial short-term variability (flickering) throughout the observations. The object is bluer in the bright state and becomes redder as its brightness decreases.
Because the spectroscopic and photometric observations were obtained at different epochs, we performed photometry using the flux-calibrated spectra obtained in 2015 in the band corresponding to the $V$ band. The result of this photometry is presented in the middle panel of Fig.~\ref{fig:color} by the red rectangles. The other photometric data folded with the orbital period were shifted to the spectrophotometric data to facilitate the comparison between the light-curve shapes. The long-term photometric behaviour presented in Appendix~\ref{sec:ap-A} shows that \rxs{} exhibits changes in its mean brightness between different observing epochs. In particular, the spectroscopic observations obtained in 2015 and the photometric observations obtained in 2017 do not correspond to exactly the same photometric state . Therefore, the comparison shown in Fig.~\ref{fig:color} is intended as a qualitative comparison of the orbital and phase-dependent behaviour, rather than as a simultaneous spectro-photometric analysis.

\subsection{The system parameters}
\label{sec:syspar}

Considering the mass--period and radius--mass relationships given by
\citet{Smith:1998}, we estimated for the secondary star in
\rxs{} a mass of $0.15\pm0.04\,\mathrm{M_{\odot}}$ and a radius of
$R_{2}=0.23\pm0.01\,\mathrm{R_{\odot}}$.
These values are also consistent with the empirical donor
sequence for cataclysmic variables at similar orbital periods reported
by \citet{Knigge:2011}. These parameters correspond to an
M5($\pm0.5$)V secondary star, with an effective temperature of
$3060^{+50}_{-90}$~K and an expected $R$-band magnitude of
approximately 20.3, which is significantly fainter than the
observed brightness at the minimum of the $R$-band light curve.
The mass of the white dwarf was estimated to be $0.75^{+0.16}_{-0.11}\,\mathrm{M_{\odot}}$ by \citet{Rawat:2023},
based on modeling the X-ray emission from the post-shock region. If this estimate is adopted, together with the secondary-star mass obtained above, a system mass ratio of $q\approx0.2$ is derived. Assuming also that the low-velocity component of the emission lines is formed close to the L$_1$ point, and adopting these masses, we estimate
\footnote{The estimate follows the approach of
\citet{1992MNRAS.257..476D}, in which the radial-velocity
semi-amplitude of the emission originating near the irradiated inner
face of the donor is used to constrain the orbital inclination.}
a system inclination of $47^{\circ}\pm10^{\circ}$.

The heating of the secondary star by high-energy radiation emitted
from the accreting magnetic white dwarf results in a shift of the emission-line formation region toward the irradiated inner hemisphere of the donor \citep{1992MNRAS.257..476D}.
Therefore, the radial-velocity semi-amplitude of this component may not directly trace the motion of the secondary star's center of mass. Consequently, both the mass ratio and inclination quoted above should be regarded as estimates that depend on the adopted model and assumptions, rather than as independent dynamical determinations of the system parameters.

\subsection{Doppler Tomography}
\label{subsec:Doppler}

\begin{figure*}
\centering
    \includegraphics[width=0.4\textwidth]{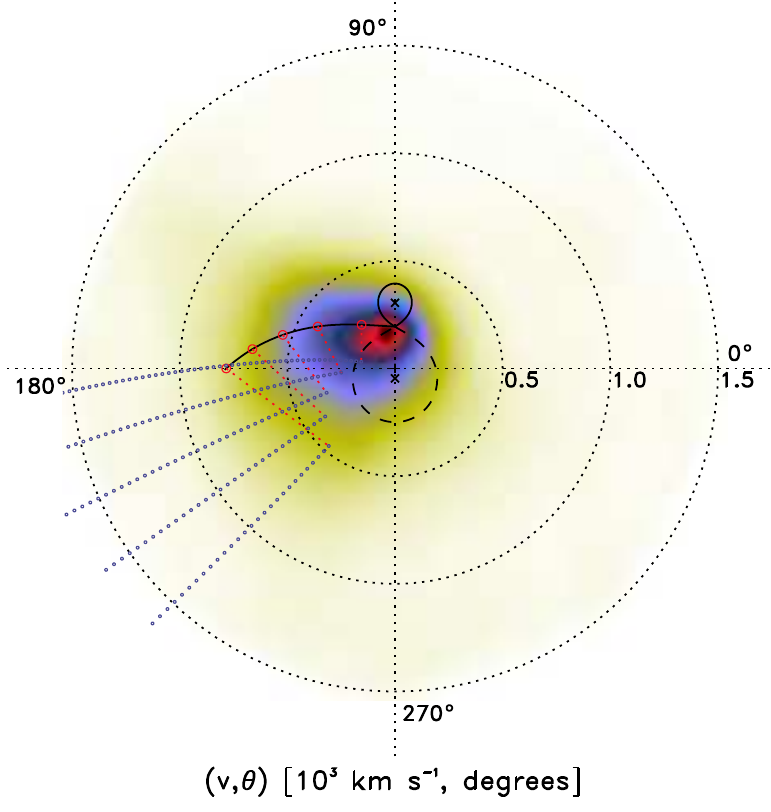}
    \hspace{0.06\textwidth}
    \includegraphics[width=0.4\textwidth]{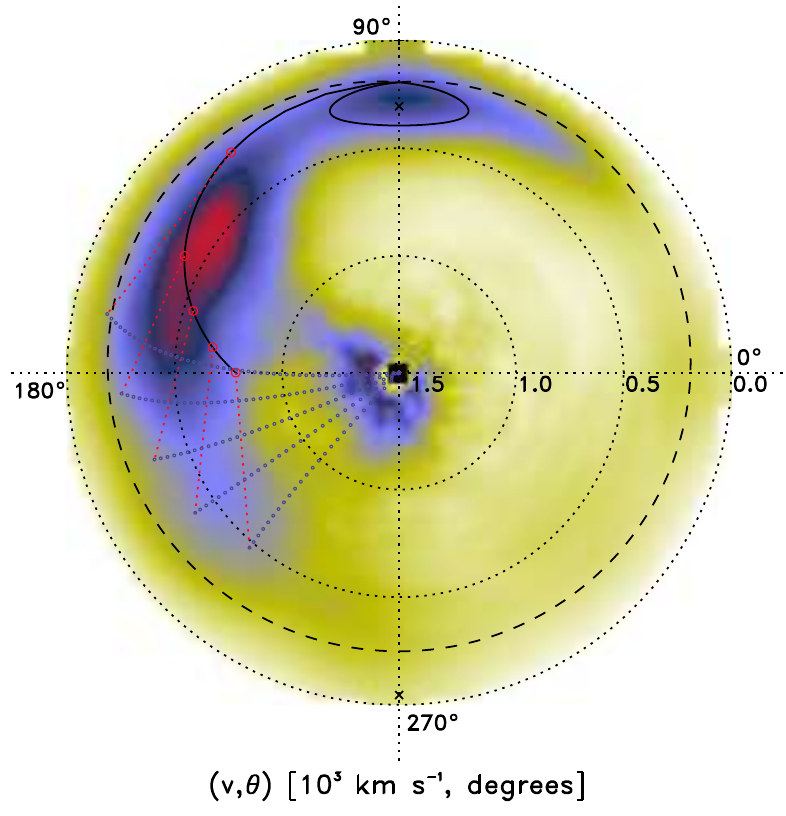}\\[0.5em]

   \includegraphics[width=0.4\textwidth]{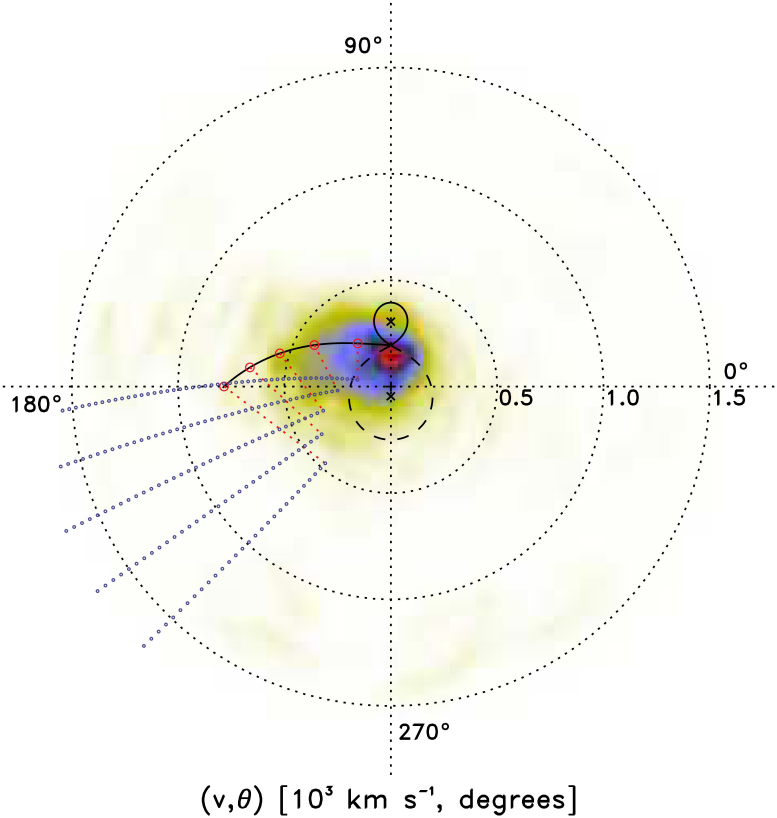}
    \hspace{0.06\textwidth}
    \includegraphics[width=0.4\textwidth]{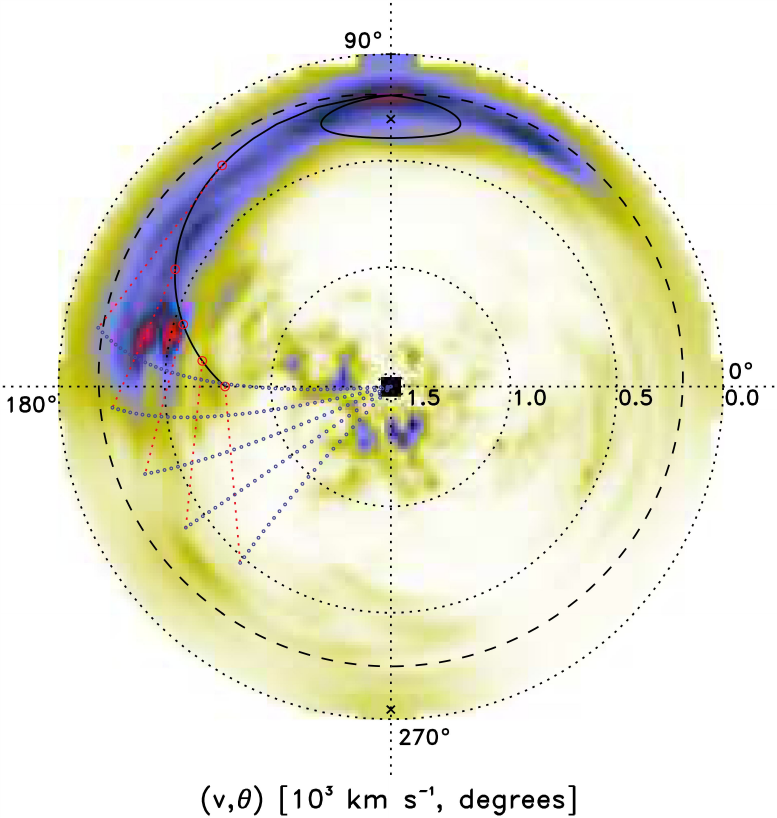}\\[0.5em]

    \includegraphics[width=0.4\textwidth]{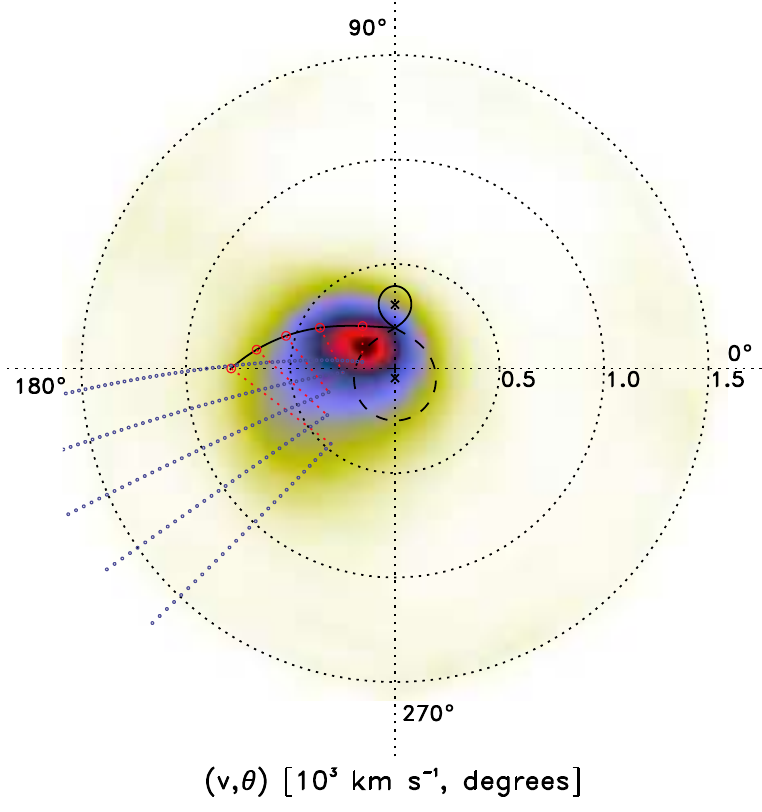}
    \hspace{0.06\textwidth}
    \includegraphics[width=0.4\textwidth]{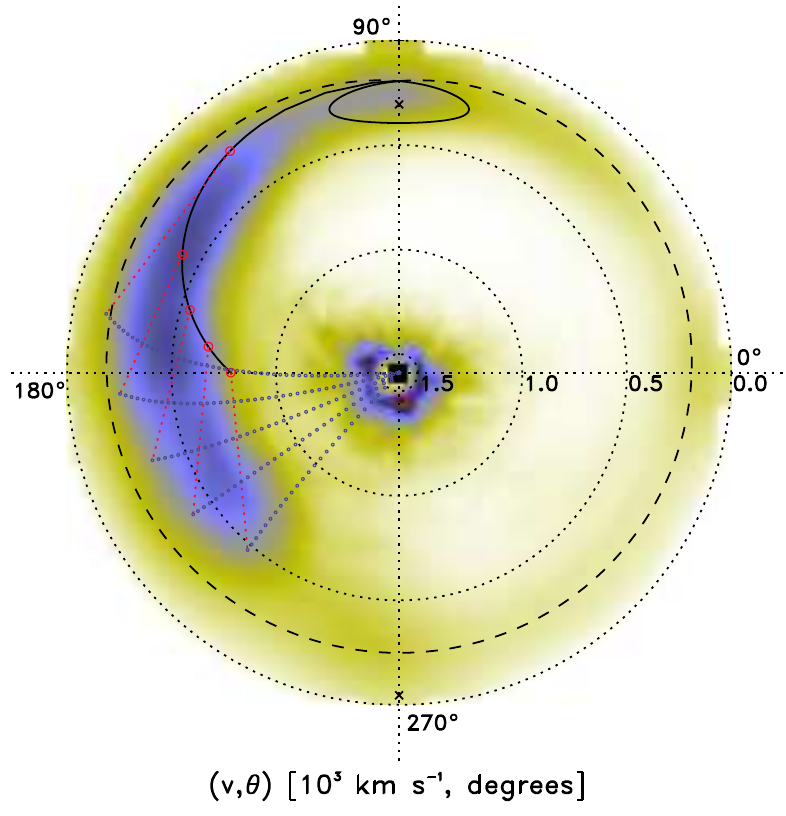}
    
    \caption{Doppler tomography of 1RXS J174320.1$-$042953. \textbf{Left column:} standard-projection Doppler maps.  \textbf{Right column:} inside-out projection maps. From top to bottom: H$\beta$, \ion{He}{i} $\lambda5015.67$, and \ion{He}{ii} $\lambda4686$. The overplotted lines indicate the ballistic accretion stream trajectory and the Roche lobe of the secondary star, computed using the system parameters derived in Section~\ref{sec:syspar}.}
  \label{fig:tomo}
\end{figure*}

To unravel the internal structure of \rxs{}, we applied Doppler tomography to our spectroscopic data. This technique, originally developed by \cite{Marsh:1988} and later refined by \cite{Kotze:2016}, transforms spectra ordered by orbital phase into two-dimensional maps in radial-velocity space ($V_{\rm x}$, $V_{\rm y}$). In particular, the inside-out projection reverses the conventional velocity mapping by placing zero velocity at the outer edge of the tomogram and higher velocities toward the center, thereby improving the visibility of high-velocity structures that may be compressed in standard Doppler maps.
The standard-projection Doppler tomograms of the H$\beta$, \ion{He}{ii} $\lambda$4686 and \ion{He}{i} $\lambda$5015.67 lines are shown in the left-hand panels of Fig.~\ref{fig:tomo}. The maps were constructed using the system parameters presented in Section~\ref{sec:syspar}. In all three cases, the emission is dominated by a single compact region in velocity space. A slight elongation of the emitting regions is also observed, which may be related to the direction of the accretion flow. No ring-like structure that could be associated with a Keplerian accretion disc is detected, reinforcing the polar nature of \rxs{}. A small phase offset between the emission centroids of the Balmer, \ion{He}{i}, and \ion{He}{ii} lines is also apparent. The Doppler tomograms obtained using the inside-out projection are shown in the right-hand panels of Fig.~\ref{fig:tomo}.

The H$\beta$ Doppler maps reveal a dominant and compact emission region in velocity space, with no evidence for a complete ring structure as expected from an accretion disc. The inside-out H$\beta$ tomogram additionally shows emission consistent with the irradiated secondary star, together with extended emission associated with the accretion stream and the accretion curtain. The detection of emission linked to the secondary in H$\beta$ suggests that this lower-excitation line traces a wider range of physical regions within the system, including cooler material located outside the immediate vicinity of the white dwarf.
The Doppler tomograms of the \ion{He}{i} $\lambda$5015.67 line show an emission distribution that is intermediate between H$\beta$ and \ion{He}{ii}. The emission remains compact and concentrated in a localized region of velocity space, with no indication of a disc-like structure. In the inside-out projection, \ion{He}{i} highlights emission associated with the accretion flow, although with less spatial extension than observed in H$\beta$. A weak contribution consistent with the irradiated secondary star may be present, suggesting that \ion{He}{i} traces moderately excited material within the accretion stream and curtain.
In contrast, the Doppler tomograms of the \ion{He}{ii} $\lambda$4686 line exhibit a more compact emission distribution in velocity space than H$\beta$ and \ion{He}{i}, consistent with an origin in hotter regions closer to the white dwarf. As in the case of H$\beta$, the standard \ion{He}{ii} tomogram shows no evidence for emission associated with an accretion disc. The inside-out \ion{He}{ii} tomogram clearly highlights emission from the accretion stream and the accretion curtain, while no significant contribution from the secondary star is detected, as expected for a higher-excitation line.

\section{Discussion}
\label{sec:discusion}

Our spectroscopic and photometric observations indicate that \rxs{} exhibits the characteristic properties of a polar-type magnetic cataclysmic variable. The optical spectra are dominated by single-peaked emission lines, with strong emission of \ion{He}{ii} comparable in intensity to H$\beta$. No evidence of cyclotron harmonics is detected in the optical continuum, suggesting either a relatively high magnetic field, for which the cyclotron emission peaks outside the optical range, or that the cyclotron emission was intrinsically weak during the observed accretion state.

High-time-resolution photometric observations reveal strong flickering in the light curve, indicative of non-homogeneous accretion onto the white dwarf. Combined with phase-resolved spectroscopy, the data show the presence of a broad eclipse of the main emitting region, approximately one magnitude deep, occurring near orbital conjunction when the secondary star lies between the white dwarf and the observer. At this phase, a redshifted absorption component with velocities up to $\sim1400\,\mathrm{km\,s^{-1}}$ is observed, indicating the presence of absorbing material along the line of sight.

To date, emission--absorption structures in optical emission lines have been reported in only a small number of polar systems, including MN Hya \citep{Ramsay1998}, FL Cet \citep{2005ApJ...620..422S}, V808 Aur \citep{2016AstBu..71..101B}, MASTER OT J132104.04+560957.8 \citep{2018AJ}, ZTF J0850+0443 \citep{Rodriguez2023}, and 1RXS J184542.4+483134 \citep{Kochkina2023}. The main properties of these systems, together with those of \rxs{}, are summarized in Table~\ref{tab:line_reversals}. Most of these systems have high inclinations and/or show eclipses, although MASTER OT J132104.04+560957.8 and \rxs{} show that this type of structure can also be observed in non-eclipsing systems.

\begin{table*}
\centering
\caption{Polar systems showing emission--absorption line reversals.}
\label{tab:line_reversals}
\begin{tabular}{lcccccl}
\hline
System & $P_{\rm orb}$ & $i$ & Eclipsing & Lines & Phase & Reference \\
       & (min)         & (deg) &           &       &       &           \\
\hline

MN Hya &
203.4 &
$\sim75$ &
Yes &
H$\alpha$ &
$\sim0.9$ &
\citet{Ramsay1998}
\\

FL Cet &
87.1 &
$\sim85$--$90$ &
Yes &
Balmer, He I &
$\sim0.8$--$0.9$ &
\citet{2005ApJ...620..422S}
\\

V808 Aur &
117.6 &
$\sim79$ &
Yes &
Balmer, He I, He II &
Pre-eclipse &
\citet{2016AstBu..71..101B}
\\

MASTER OT J132104.04+560957.8 &
91 &
--- &
No &
Balmer, He I, He II &
$\sim0.95$--$1.06$ &
\citet{Littlefield2018}
\\

ZTF J0850+0443 &
103.2 &
High &
Yes &
Balmer, He I &
Pre-eclipse &
\citet{Rodriguez2023}
\\

1RXS J184542.4+483134 &
$\sim79$ &
$79.7$--$84.3$ &
Yes &
Balmer, He I &
Pre-eclipse &
\citet{Kochkina2023}
\\

\rxs{} &
124.4 &
$\sim47$\footnotemark &
No &
H$\beta$, He II &
$\sim0.0$ &
This work
\\

\hline
\end{tabular}

\footnotetext{Indicative inclination obtained under the assumptions
described in Section~\ref{sec:syspar}.}
\end{table*}

MN Hya is an eclipsing polar with an orbital period of $P_{\rm orb}=203.4$ min and an inclination of $i\approx75^\circ$. Prior to eclipse, its emission lines weaken and complex emission--absorption profiles appear, with absorption components redshifted by up to $\sim600$ km s$^{-1}$ relative to the emission peaks. In FL Cet ($P_{\rm orb}=87.13$ min, $85^\circ\lesssim i\lesssim90^\circ$), strong P-Cygni-type absorption develops near $\phi\approx0.88$, temporarily transforming the Balmer and neutral helium lines into absorption. V808 Aur ($P_{\rm orb}=117.6$ min, $i\approx79^\circ$) also shows absorption structures over a broad phase interval, with the absorption moving from the red to the blue wings of the Balmer lines. In the non-eclipsing system MASTER OT J132104.04+560957.8, the absorption structures progressively replace the emission profiles before the lines return to emission. In ZTF J0850+0443 and 1RXS J184542.4+483134, both high-inclination systems, the line reversals also appear at phases close to eclipse and have been associated with accretion-stream material crossing in front of the main emitting regions. These cases show that the appearance and evolution of the absorption components are closely related to the system geometry and to the orbital phase at which the accretion-flow material crosses the line of sight.

In contrast, \rxs{} appears to be a polar with a moderate inclination,
consistent with the estimate of $i=47^\circ\pm10^\circ$ discussed in Section~\ref{sec:syspar}. In this geometry, the redshifted absorption observed in \rxs{} is relatively weak and confined to a narrow orbital phase interval around $\phi\approx1.0$, without significant velocity drift across the line profiles. This behaviour can be naturally explained if the line of sight briefly intersects the accretion stream or curtain near conjunction, rather than crossing a
substantial portion of the flow as in high-inclination eclipsing
systems. A similar interpretation was proposed by \citet{Littlefield2018} for MASTER OT J132104.04+560957.8, where the
line reversals were associated with material in the accretion curtain
crossing the line of sight, and the H, He I, and part of the He II
emission were found to originate within the curtain. The presence of
this type of behaviour also in non-eclipsing systems suggests that a
high inclination favours the detection of these structures, but is not
a necessary condition for observing line-profile reversals.
 

\section{Conclusions}
\label{sec:conclusion}

We have carried out a detailed spectroscopic and multi-band photometric study of the magnetic cataclysmic variable \rxs{}. Our main results can be summarized as follows:

\begin{enumerate}
    \item
    The optical spectra are dominated by single-peaked and highly variable Balmer, \ion{He}{i}, and \ion{He}{ii} emission lines, consistent with a polar-type magnetic CV. The absence of any disc-like signature in the Doppler tomograms supports the classification of the system as a polar.
    
    \item The H$\beta$ and \ion{He}{ii} ($\lambda4686$\,\AA) emission lines can be successfully deconvolved into two Gaussian components. The low-velocity components have semi-amplitudes of $\sim170$~km~s$^{-1}$ (H$\beta$) and $\sim147$ km s$^{-1}$ (\ion{He}{ii}), while the high-velocity components reach $\sim317$ km s$^{-1}$ and $\sim460$ km s$^{-1}$, respectively.
    
    \item Doppler tomography reveals a dominant compact emission region in velocity space for H$\beta$, \ion{He}{i} $\lambda5015$, and \ion{He}{ii} $\lambda4686$, with no evidence of a Keplerian accretion disc. The inside-out projections highlight emission associated with the accretion stream and curtain, while H$\beta$ shows additional contribution consistent with irradiation of the secondary star.
    
    \item A redshifted absorption component is detected near the H$\beta$ emission line at specific orbital phases, reaching velocities up to $\sim1400$ km s$^{-1}$. This feature is likely produced by accretion-stream material crossing the line of sight. Comparison with other polars showing emission--absorption line reversals suggests that the visibility of these structures is strongly influenced by the accretion-flow geometry and orbital phase.
    
    \item Multi-band photometry confirms a periodic modulation with an orbital period of $P_{\rm orb}=0.0864 \pm 0.0001$ days. The light curves show strong flickering and a broad minimum consistent with orbital modulation in a magnetic accretion geometry.
    
    \item Using standard mass--period relations, we estimate a secondary mass of $\sim0.15\,\mathrm{M_\odot}$. If the previously estimated white dwarf mass of $\sim0.75\,\mathrm{M_\odot}$ is adopted, we obtain indicative values of $q\sim0.2$ and a system inclination of approximately $47^\circ\pm10^\circ$.

\end{enumerate}

Altogether, the spectroscopic, photometric, and tomographic properties consistently indicate that \rxs{} is a polar system in which the accretion flow is magnetically channeled onto the white dwarf without the formation of an accretion disc.

\section{FACILITIES}

This work is based on observations carried out at the Observatorio Astronómico Nacional at San Pedro Mártir (OAN-SPM), Mexico. 

\renewcommand{\refname}{REFERENCES}
\bibliography{RMxAA}


\section{ACKNOWLEDGEMENTS}

DHGB acknowledges financial support from CONACYT through grant \#319800 and from the Investigadores por M\'exico program. 
We also sincerely thank the staff and workers of the Observatorio Astron\'omico Nacional at San Pedro M\'artir (OAN-SPM) for their invaluable technical support and dedication, which made the observations possible.
Ma.T.G-D gratefully acknowledgesthe support from UNAM DGAPA PAPIIT through GrantAG101223
SVZh acknowledges DGAPA-PAPIIT grant  IN105826. 
AC acknowledges DGAPA-PAPIIT grant TA100625 and the Royal Society Newton International Fellowship NF170803
CAN acknowledges financial support from DGAPA-PAPIIT (grants IN119123 and IN112226) and SECIHTI (project CBF2023-2024-1418).
We gratefully acknowledge the support and assistance of the OAN-SPM staff during the observations.

\begin{appendices}

\section{Long-term photometric behaviour of \rxs{}}
\label{sec:ap-A}

\begin{figure*}
    \centering
    \includegraphics[width=\textwidth]{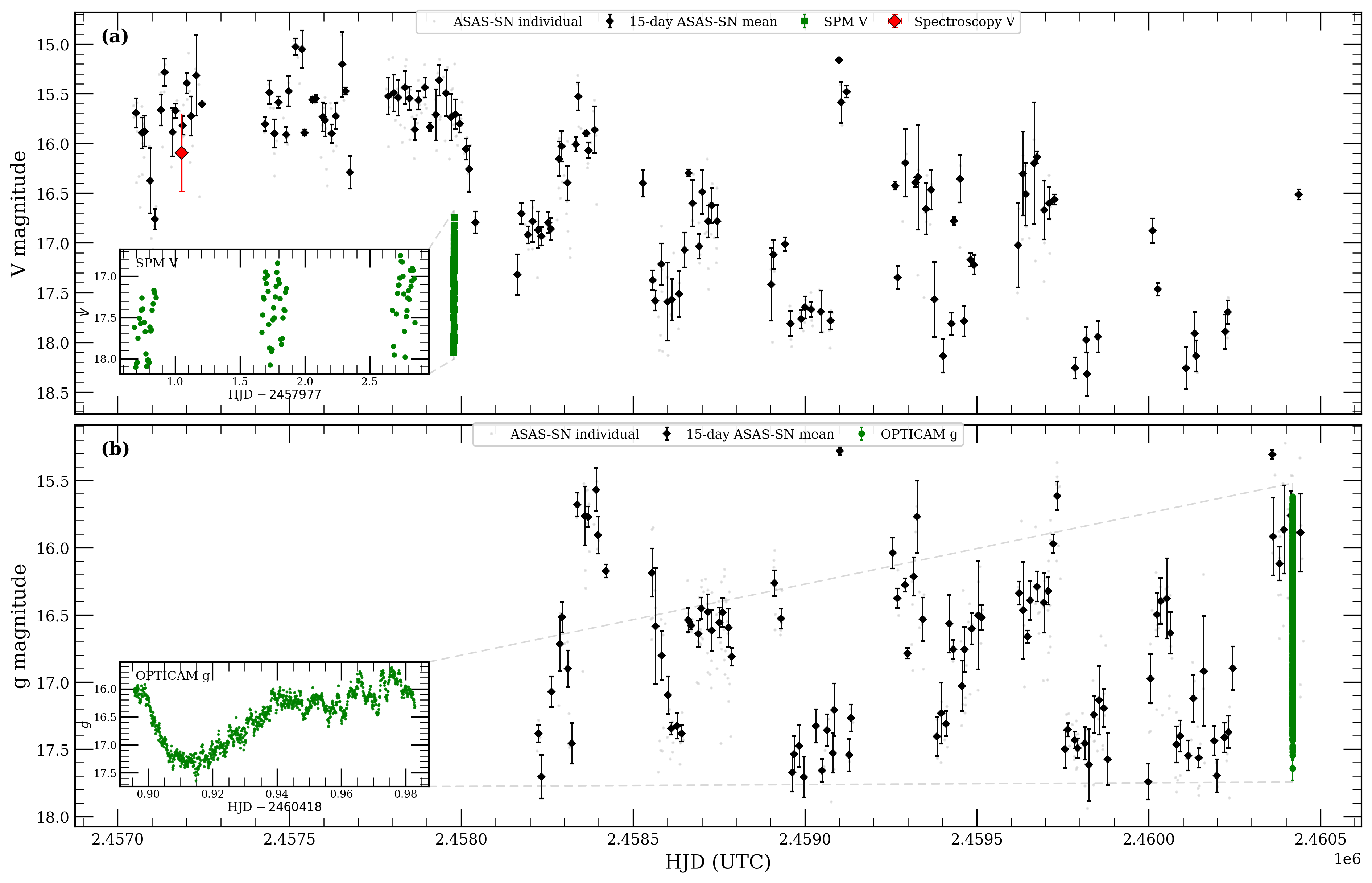}
    \caption{
    Long-term photometric behaviour of \rxs{} based on archival
    ASAS-SN photometry and our observational data.
    Panel (a) shows the $V$-band light curve, including the individual
    ASAS-SN measurements (gray points), the 15-day ASAS-SN mean
    magnitudes (black diamonds), the 2017 SPM $V$-band photometry
    (green points), and the $V$-band magnitude obtained from the
    flux-calibrated spectra acquired during the 2015 spectroscopic
    campaign (red diamond). Panel (b) shows the corresponding
    $g$-band long-term light curve, including the archival ASAS-SN
    photometry and the 2024 OPTICAM observations (green points).
    The insets show enlarged views of the SPM and OPTICAM campaigns.
    The long-term variability indicates that the 2015 and 2017
    observations were not obtained in exactly the same photometric
    state.
    }
    \label{fig:longterm_lc}
\end{figure*}

Figure~\ref{fig:longterm_lc} presents the long-term photometric behaviour of \rxs{} using archival ASAS-SN observations together with our SPM and OPTICAM data. The upper panel shows the $V$-band light curve, while the lower panel presents the corresponding behaviour in the $g$ band.
The light curves show that \rxs{} exhibits significant variability on timescales ranging from days to years. In the $V$ band, the system varies approximately between magnitudes 15 and 18.5, with changes in the mean brightness level between different observing seasons. The individual ASAS-SN measurements also show short-term variability, while the 15-day averages highlight the longer-term changes in the brightness state of the system.
The $V$-band photometry obtained at SPM in 2017 shows a clear orbital modulation, with changes in amplitude and mean brightness during the three observing nights. For comparison, we also include the $V$-band magnitude estimated from the flux-calibrated spectra obtained in 2015.
In the $g$ band, the 2024 OPTICAM observations show a pronounced decrease in brightness followed by a gradual recovery, together with strong short-term variability. These characteristics are consistent with the photometric behaviour described in Section~\ref{sec:lc}, where the system shows orbital
modulation, flickering, and brightness changes between different
epochs.

\end{appendices}

\end{document}